\documentclass[acm small]{acmart}

\usepackage{booktabs} 
\usepackage[english]{babel} 
\usepackage[utf8x]{inputenc} 
\usepackage[colorinlistoftodos]{todonotes} 

\setcopyright{rightsretained}

\begin{document}
\title{Older Adults and Crowdsourcing: Android TV App for Evaluating TEDx Subtitle Quality}


\author{Kinga Skorupska}
\affiliation{%
  \institution{Polish-Japanese Academy of Information Technology}
  \streetaddress{86 Koszykowa str.}
  \postcode{02-008}
  \city{Warsaw}
  \country{Poland}
}
\email{kinga.skorupska@pja.edu.pl}

  \author{Manuel N\'{u}\~{n}ez}
\affiliation{%
  \institution{Polish-Japanese Academy of Information Technology}
  \streetaddress{86 Koszykowa str.}
  \postcode{02-008}
  \city{Warsaw}
  \country{Poland}
}
\email{manuel.nunez@pja.edu.pl}

\author{Wies\l{}aw Kope\'{c}}
\orcid{0000-0001-9132-4171}
\affiliation{%
  \institution{Polish-Japanese Academy of Information Technology}
  \streetaddress{86 Koszykowa str.}
  \postcode{02-008}
  \city{Warsaw}
  \country{Poland}
}
\email{kopec@pja.edu.pl}

\author{Radoslaw Nielek}
\affiliation{%
  \institution{Polish-Japanese Academy of Information Technology}
  \streetaddress{86 Koszykowa str.}
  \postcode{02-008}
  \city{Warsaw}
  \country{Poland}
}
\email{nielek@pja.edu.pl}

\renewcommand{\shortauthors}{K. Skorupska et al.}

\begin{abstract}
In this paper we describe the insights from an exploratory qualitative pilot study testing the feasibility of a solution that would encourage older adults to participate in online crowdsourcing tasks in a non-computer scenario. Therefore, we developed an Android TV application using Amara API to retrieve subtitles for TEDx talks which allows the participants to detect and categorize errors to support the quality of the translation and transcription processes. It relies on the older adults' innate skills as long-time native language users and the motivating factors of this socially and personally beneficial task. The study allowed us to verify the underlying concept of using Smart TVs as interfaces for crowdsourcing, as well as possible barriers, including the interface, configuration issues, topics and the process itself. We have also assessed the older adults' interaction and engagement with this TV-enabled online crowdsourcing task and we are convinced that the design of our setup addresses some key barriers to crowdsourcing by older adults. It also validates avenues for further research in this area focused on such considerations as autonomy and freedom of choice, familiarity, physical and cognitive comfort as well as building confidence and the edutainment value. 
\end{abstract}


\setcopyright{acmlicensed}
\acmJournal{PACMHCI}
\acmYear{2018} 
\acmVolume{2} 
\acmNumber{CSCW} 
\acmArticle{159} 
\acmMonth{11} 
\acmPrice{15.00}
\acmDOI{10.1145/3274428}

\received{April 2018} 
\received[revised]{July 2018}
\received[accepted]{September 2018}



\maketitle

\section{Introduction}

According to the latest population data published by Eurostat the share of older adults, defined as people aged 65+, is increasing in every member and candidate state. In European Union member states (EU-28) this share has risen by 2.4\% between 2006 and 2016. Moreover, the long term 2015 EUROPOP projection covering the time up to 2080 shows that this trend will continue, and people aged 65+ are expected to comprise 29.1\% of the total EU-28 population by 2080, while latest data shows they already made up 19.2\% of the population in 2016 \cite{population_structure2017}. The same is increasingly true of Western societies all over the world \cite{UNreport}. For example, according to the U.S. Census Bureau by the year 2050, the population aged 65+ in the United States will almost double, reaching over 20\% of the society \cite{ortman2014aging}. 
While the number of older adults is already significant, their potential remains largely untapped because of a shortage of adequate research-informed activities and programs allowing them to contribute to the society. Moreover, numerous studies confirm that developing sustainable solutions for older adults is still challenging. In particular, Knowles and Hanson underscore that there is still room for further research and establishing a more holistic approach to the problem, despite extensive progress in that field over the last 20 years \cite{knowles2018wisdom,knowles2018older}. Access to a lot of available activities relies heavily on ICT skills and familiarity with computers and technology, as well as the accessibility of the solutions. Moreover, older adults are often not provided with adequate motivation to take part in these activities, e.g. crowdsourcing \cite{brewer2016would}, as they are not challenging, fun or do not hold an explicit connection to real life. 

It is our intention to address some of these concerns by designing and testing the feasibility of an engaging crowdsourcing solution for older adults that would encourage them to participate in crowdsourcing tasks. Therefore, we decided to assess older adults' interaction and engagement with an online crowdsourcing task relying on their innate language skills, at home in a non-computer scenario. For this purpose we launched a qualitative pilot study to build and test a language-focused solution and verify the underlying concept and barriers, including the interface, configuration issues, interaction, topics and the process itself.

In the course of the study, we developed and tested an Android TV application for crowdsourcing TED and TEDx subtitle errors that enables the participants to detect and categorize them. In this study we operate at the intersection of interests of four groups of potential stakeholders, depicted in Table \ref{stakeholders}. 

\begin{table}
\caption{Four groups of stakeholders}
\label{stakeholders}
\begin{tabular}{p{2.5cm}p{10cm}}
	\hline
	\textbf{Group} & \textbf{Their stake} \\
	\hline
    Older adults & as the target group for our solution and the crowdsourcers motivated by the edutainment aspect of the task and its positive impact \\
    \hline
   Translation communities 
   & as direct beneficiaries of the insights from the errors detected and problems signaled by the crowdsourcers\\
	\hline
   General public & as the ones benefiting from the improved quality of subtitles \\
   	\hline
   Developers & as the ones involved in addressing the barriers to crowdsourcing and benefiting from insights from our development process and tests\\
   	\hline
\end{tabular}
\end{table}

Our assumptions, based on our previous work with older adults in a similar field \cite{kopec2017living}, 
were that they enjoy using Android tablets \cite{kopec2017location} and therefore can learn to proficiently use Smart TVs, which have the benefits of relying on a familiar setting (TV set) and interfaces (e.g. Teletext), a simple remote, as well as sport a larger screen size. Furthermore, older adults are experienced users of their native language and can ascertain if utterances seem natural. 
It was our expectation that they would see improving the quality of subtitles for TED and TEDx talks as an interesting and motivating task due to its educational value.

Therefore, our study aims to explore a few research aspects. Our prime objective was to determine the potential of Smart TVs as crowdsourcing platforms for older adults. This necessitated to first answer the question of whether older adults interact with Smart TVs in a more familiar fashion than with computers. Secondly, we wanted to ascertain if they consider themselves competent enough in their native language to detect errors in subtitles. And lastly, we examined if such a task is engaging for them, and what are the emergent barriers to their involvement, and possible ways to overcome them.
We hope that researching these questions will help us draw conclusions on how to better include older adults in ICT-dependent crowdsourcing tasks, and in online activities with the use of some new, but familiar interfaces and to find promising avenues for further study. At the same time, we want to propose and verify the design and setup of online crowdsourcing translation process support based on commodity goods, in this case the Smart TV platform, and to mitigate the barriers that may exist for this type of activity, also related to vital challenges of crowd work systems raised at CSCW in recent years \cite{kittur2013future,culbertson2017have,zhu2014reviewing,yu2014comparison}.

The paper is organized as follows: first we present the related works on topics, such as older adults' empowerment, crowdsourcing and volunteering, subtitling TED and TEDx, designing for older adults and Android TV in particular; second, we describe our methods, including the design of the DreamTV crowdsourcing tool and the testing protocol; third the results section follows, providing a summary of our collected qualitative data; finally, we interpret our results in the discussion section, which is followed by our conclusions and future research plans.

\section{Related work}

\subsection{Challenges of empowering older adults}

There exist increasingly more studies at the intersection of HCI and aging, but as Vines et al. concluded in their discourse analysis, many focus on stereotypes related to health, socialization and technology \cite{articleageoldproblem} instead of exploring the aging process and looking for opportunities. One positive example is Carrol et al.'s study of older adults as "organizational firekeepers" \cite{articlefirekeepers} in roles as keepers of history, co-designers and members of intergenerational teams contributing valuable complementary skills. So, while it is true that some older adults do not have faith and confidence in their skills, especially if they are ICT related \cite{sandhu2013ict}, such barriers can be partially mitigated to enable their valuable contributions. One way to do this is by introducing a positive social context and support \cite{orzeszek2017design}. While in our previous research this context was strongly based on direct social interaction between generations \cite{kopec2017location,kopec2017older}, in our current study it is built on the clear social benefit of the opportunity to crowdsource edutainment content \cite{rapeepisarn2006edutainment}: having a lasting positive effect on a large number of viewers using subtitles e.g. for learning languages and educating the general population. This translates into a strong positive motivator, as one of the most common needs expressed by older adults is the need to feel useful, help others and contribute to the common good in a social setting \cite{brewer2016would}. 
Additionally, it also uses one of the advantages that older adults hold over the younger generation: their experience, explored for example by Balcerzak et al. \cite{balcerzak2017F1,balcerzak2017close}. Just as their vast knowledge of the cultural and historical context had a positive impact on their performance in a historical location based game \cite{kopec2017location}, so can their long time experience with their native language act as an empowering factor, allowing them to feel confident, and competent enough to climb the ladder of ICT proficiency (a benefit they are aware of \cite{aula_learning_2004}); eventually even making them ready to join the ICT solution development process, as in the SPIRAL method \cite{kopec2017spiral}.
On top of the question of feeling competent, there are specific physiological changes in the brain of older adults, which comprise of deterioration of working memory, and consequently, the ability to acquire new information \cite{wolfson2014older}. 
But, this effect can be mitigated by creating step-by-step instructions for doing an ICT task in a workshop format \cite{kopec2017living}, and specific instruction design elements targeting these problems \cite{wolfson2014older}. At the same time, there is evidence that crystallized intelligence, which consists of general knowledge derived from experience, not only does not decline \cite{wang1993changes}, but may benefit from aging \cite{mcardle2000modeling}. This may prove to be an asset in tasks related to the native language, e.g. subtitling, as crystallized intelligence is primarily measured through general world knowledge and language tasks \cite{baghaei2015c}. 

\subsection{Volunteering and crowdsourcing}

Encouraging older adults' engagement in volunteering and crowdsourcing activities in general \cite{morrow2010volunteering} benefits the society as a whole. Such activities have positive effects on older adults' well-being \cite{morrow2003effects}, as well as their mental and physical health \cite{lum2005effects}. Thus, they can be regarded as a protective factor for their psychological well-being \cite{greenfield2004formal,hao2008productive}. In particular, they can also pose certain health advantages, as staying active and learning new things can delay the onset of age-related issues, e.g. mild cognitive decline and some related aspects of dementia \cite{kotteritzsch2014adaptive}. These benefits were confirmed in a study of Lum and Lightfoot, who showed that volunteering slows the negative effects of aging and helps to combat depression \cite{lum2005effects}. They may extend to online crowdsourcing tasks, as the mental benefits may be connected to the results of the study by Yang and Cheng-Yu \cite{yang2010motivations}, who found that Wikipedia editors are internally driven by self-concept motivation: their need to maintain a consistent and positive image of themselves, which they thus satisfy. This issue was also explored in previous studies on Wikipedia editing by older adults \cite{nielek2017wikipedia}. As such, engaging older adults in interesting tasks which have a high social value may be beneficial on all fronts. 

Older adults have proven to be more aware of their needs and abilities than the younger generation \cite{djoub_ict_2013}; therefore, they are more selective in choosing their activities. Moreover, they differ from the younger generation in their online behavior and decision-making \cite{von2018influence}, which, alongside their generally lower ICT skills, may explain how little interest they expressed in the Mechanical Turk platform populated by tedious and repetitive crowdsourcing tasks \cite{brewer2016would}. 

Volunteering and user engagement are often considered in terms of using gamification techniques, i.e. the use of game design elements in non-game contexts \cite{deterding2011game}. Some researchers claim that gamification is a more psychological than technological issue \cite{zichermann2011gamification}. Many methods and gamification tools are based on sound psychological foundations, like the well-established self-determination theory \cite{ryan2000self}, which was verified in the context of older adults \cite{vallerand1989motivation}. Although there are some promising reports on using gamification elements in older adults crowdsourcing tasks \cite{itoko2014involving}, there are also more recent reports that show difficulties in this field \cite{brewer2016would}.

\subsection{Subtitling TED and TEDx talks}

TED\footnote{When TED first launched it was an invite-only conference combining Technology, Entertainment and Design (TED) which eventually became an annual event in California. In 2001 the conference was acquired by Chris Anderson, who in 2006 shared the recorded talks online, paving the way for the current popularity of TED as an edutainment platform. More information about TED is available online at: https://www.ted.com/about/our-organization/history-of-ted} is a nonprofit organization that organizes conferences on the topics of technology, entertainment and design. Its mission is to share "ideas worth spreading" with the wider audience through the talks recorded during its conferences. To popularize the format across the world, TED launched the TEDx\footnote{The TEDx programme allows volunteers to apply for a TEDx license to organize a TED-like conference in their local area. More information about TEDx is available online at: https://www.ted.com/about/programs-initiatives/tedx-program} initiative of volunteer organized TED-like conferences around the globe. Since 2009, when this initiative was launched, the number of published TEDx talks in over 100 different languages has reached 100,000. These talks totaled a billion views in large thanks to the TED Open Translation Program, now rebranded as TED Translators\footnote{The TED Translators programme allows volunteers to create subtitles for TED, TEDx and TED-Ed talks on Amara. More information about this programme is available online at: https://www.ted.com/about/programs-initiatives/ted-translators}. Within the TED Translators initiative volunteers contribute subtitles to TED, TEDx and TED-Ed videos in 100+ languages, making them accessible to everyone.  

The TED Translators initiative is a true example of a \textit{community of practice} which Leave and Wenger \cite{lave_wenger_1991} describe as a group driven by common interest that collectively improves and gains knowledge. In 2014, Cámara, in a survey-based study \cite{camara2014multilingual}, identified the need to contribute to the TED mission of "spreading ideas" as a key volunteering motivation of the TED Translators. As such, improving the quality of these subtitles is a promising example of a potentially engaging task, which is beneficial both socially (granting access to all) and individually (learning about interesting topics).
As the subtitles are sourced in a large part from non-professional volunteers, who make mistakes that can be detected by experienced native speakers of the language, this makes the task doubly important and rewarding.

Additionally, within the TED Translators project, there exist various challenges, largely dependent on the language community. For example, not enough volunteers to satisfy the ever-growing demand for translations; as of April 17, 2018 there are 56,617 TEDx videos added to the TED team in Amara\footnote{Amara is a subtitling platform where people can volunteer to create subtitles for videos with the use of a streamlined online editor. More on Amara is available online at: https://amara.org} \cite{jansen2014amara}, an in-browser subtitling platform used by TED for its TED Translators project. Moreover, as TED Translators employ a 3-step quality assurance process, there are not enough experienced reviewers (step 2) and Language Coordinators (step 3) who can correct, and approve the subtitles for the talks so that they can be published.  

Thus, not only is there potential to introduce a fourth Quality Assurance step, or to facilitate steps 2 and 3, but also to employ Machine Transcription and Translation to grant the international audience access to the content of the videos that were not yet translated. The inclusion of Machine Translation (MT) in the human translation process, especially where quantity is important, is a clear path ahead as it can shorten the process by up to 40\% \cite{etchegoyhen2014machine}. This trend is also evident in the development avenues of market-leading Computer Assisted Translation (CAT) tools, such as MemoQ or Trados. There is clearly room for involving crowdsourcing and citizen science in the context of Machine Learning and Natural Language Processing of the subtitles e.g. through supervised learning processes. Increasing both the quantity and the quality of subtitles is important, because they provide access to a wide range of ideas, also scientific, to the hard of hearing, non-native speakers and the general public in an easy to digest, edutainment format. So, making them more accessible can be a positive motivator for the volunteers. 

\subsection{Designing for older adults and Android TV}

Although older adults are a very non-uniform group in terms of their ICT-skills and needs, there are some general guidelines, also relevant in Smart TV solution design. In "Designing Displays for Older Adults", Pak noted the older adults' need for context, consistency and low working memory burden, and their low tolerance for design clutter \cite{pak_mclaughlin_2011}. A similar focus on the need to simplify is visible in the findings of the study by Silva et al. \cite{articledontfall} concerning a TV dance application, where users likely experienced an information overload due too much information being visible on the screen when trying to repeat dance moves. More generally, Pan et al. \cite{pan2015effects} proposed a framework which consists of "symbolic familiarity, cultural familiarity, and actionable familiarity", all of which have positive effects on the adoption of technology by older adults. 
Also, as shown by our research involving an Android location-based game, \cite{kopec2017location} older adults are quite comfortable with tablets. According to Tsai et al. \cite{tsai2015getting}, this is due to the larger screen size than phones. Thus, as large screen size and design familiarity, consistency and simplicity are key considerations, for our exploratory study we decided to make use of the increasingly popular Smart TVs. Moreover, such TV sets can be controlled by a remote, which may mitigate older adults' possible problems with "mapping actions to devices" signaled by Fisk et al. \cite{fiskdesigningolderr}. So through focusing on Smart TVs, we not only deliver a large screen size, but also a familiar and simple experience of using a remote control and interacting with text on screen, not unlike Teletext. Some preliminary studies on using Smart TVs for older adults in the context of Living Labs had been initiated \cite{alaoui_lewkowicz_2015}, however, this is still a largely unexplored area of research with few general studies on interfaces for TV apps for older adults, such as the one by Nunes et al., who among other things, recommended minimizing navigation steps, clearly marking selections, using the middle of the screen, simple language and allowing for enough time to read \cite{articlereccstv}.  This encouraged us, based on our findings concerning older adults and Living Labs \cite{kopec2017living}, to explore the new technological, but familiar, opportunities that Smart TVs present in the home environment, especially in the context of crowdsourcing.
According to numerous analyses, there is still tremendous dominance of using TV sets over computer and mobile equipment, especially among older adults (defined as 65+). For example, Nielsen research on the US market shows that older adults spend over 50 hours a week in front of the TV, while fewer than 5 hours in front of the PC, including using video on PC (less than an hour a week).\footnote{More information on TV viewing habits by age can be found online at: https://www.recode.net/2016/6/27/12041028/tv-hours-per-week-nielsen} 
Although there are numerous works on various aspects of designing services and interfaces for older adults \cite{dickinson2007methods}, in particular for interactive television use-case scenario \cite{carmichael1999style}, the development of new trends of Smart TV, VOD (Video On Demand) and OTT (Over the Top) internet-based streaming services and applications, which all largely allow users to freely choose content to watch, led us to thoroughly rethink the service we planned in this crowdsourcing use case with older adults.

\section{Methods}

Based on our research and experience, we developed a pilot of an exploratory study concept for using crowdsourcing without the computer in a TV-enabled setup, as well as the tool to support the subtitle translation and review processes: the DreamTV application. 
The findings from our previous work in this field in the Living Lab context \cite{kopec2017living,nielek2017emotions,orzeszek2017design}
informed the design of our testing protocol, which presents subtitling as a task useful for a wide group of viewers (motivation), and introduces the users to its basic concepts, as well as verifies their understanding with an on-paper exercise (empowerment). We believe that the design of our study addresses key barriers to information technology and crowdsourcing by older adults, as presented in Table \ref{barriers}.  

\begin{figure}
\centering
   \includegraphics[width=0.5\textwidth]{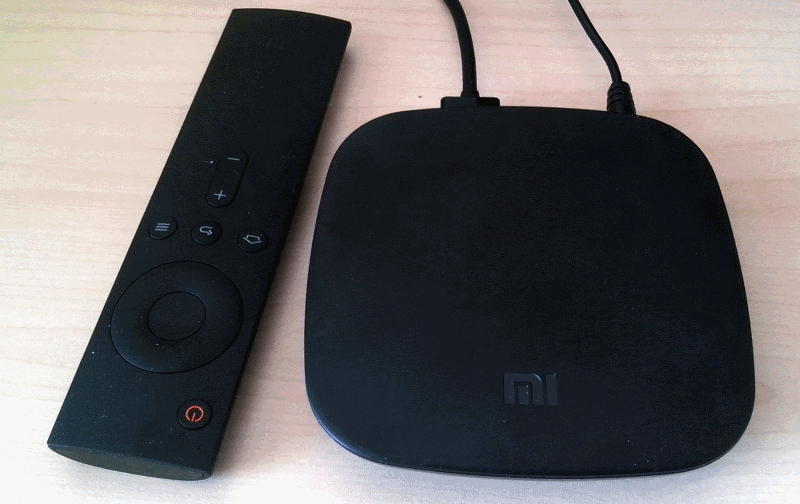}
   \caption{The Xiaomi MiBox, a TV Set-top box (STB) with a simple remote can turn any TV with an HDMI socket into a Smart TV with the Android TV OS. (Source: Wikipedia, CC BY-SA 3.0 license)}
   \label{fig:mibox}
\end{figure}

\begin{table}
\caption{Overview of signaled barriers to crowdsourcing by older adults with proposed solutions inspired by literature.}
\label{barriers}
\begin{tabular}{ll}
	\hline
	\textbf{Barriers} & \textbf{Solutions} \\
	\hline
    Uncomfortable and costly setup \cite{kopec2017location,sandhu2013ict}& TV sets at home \cite{alaoui_lewkowicz_2015,kopec2017living} \\
    \hline
   Unfamiliar interfaces \cite{orzeszek2017design,aula_learning_2004,czaja2006factors,czaja2009information} & Text and remote control \cite{fiskdesigningolderr}\\
	\hline
   Repetitive tasks \cite{brewer2016would} & Educational wide-domain TEDx videos \cite{thenewyorktimes_2017}\\
   	\hline
   Unclear social benefit \cite{brewer2016would} & Improving subtitles for all \cite{camara2014multilingual}\\
   	\hline
   Unclear personal benefit \cite{brewer2016would,articleageoldproblem} & Learning, positive self-image \cite{yang2010motivations,nielek2017wikipedia}\\
   	\hline
   Unsocial nature of the task \cite{vines2011eighty} & Part of the community \cite{articlefirekeepers,nielek2017emotions} \\
	\hline
\end{tabular}
\end{table}

Here, it is important to note that subtitles have only started to become more common in Poland in the 1990s, mostly in cinemas and later online. The preferred TV format is using a voice-over translation, which is also a popular practice in Cambodia, Mongolia, Vietnam and some other East European countries.\footnote{Voice-over translation is common in part as it is cheaper than dubbing, as it usually involves only one voice actor, and there is no need to mix the original soundtrack which just plays quiet in the background, but also because of force of habit.} Unlike dubbing, which employs a cast of actors to recreate the soundtrack in the target language, voice-over translation usually employs the technique of using a single voice actor talking over a quieter original film soundtrack. This fact may form an additional barrier for the older adults' engagement, as our tests were conducted with Polish participants using Polish language subtitles.

\subsection{DreamTV application}

Therefore, based on our analysis of possible solutions to the aforementioned barriers and taking into account the needs of various translation communities, for the purpose of our exploratory study we developed the DreamTV application, which allows users to detect errors in subtitles. Its main features are described below. 

\subsubsection {Video selection}

\begin{figure}
\centering
\includegraphics[width=0.5\textwidth]{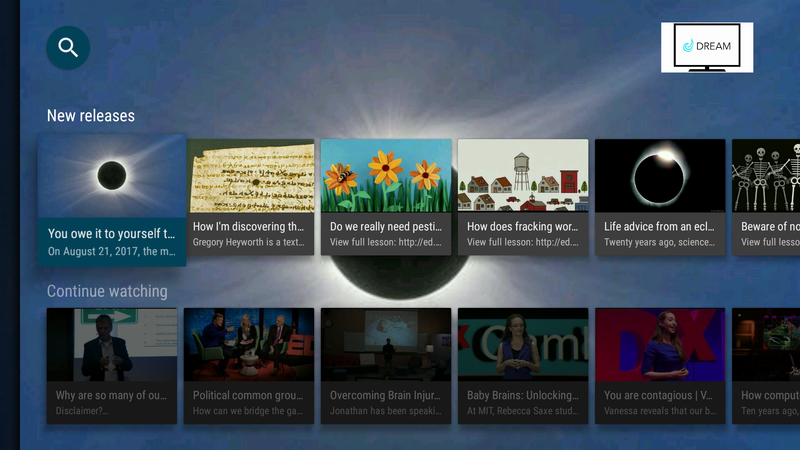}
\caption{The main screen of DreamTV allows the users to select new videos, or continue watching their previous ones.}
\label{fig:main_screen} 
\end{figure}

The application allows the users to choose the videos to play (Fig. \ref{fig:main_screen}). It remembers the previous choices and the progress of the user on each video, making it possible to resume the tasks. It keeps track of the users' contributions and generates statistics. The identified errors are written in a database, which is accessible online.

\subsubsection {Playing the video mode}
\begin{figure}
\centering
   \includegraphics[width=0.5\textwidth]{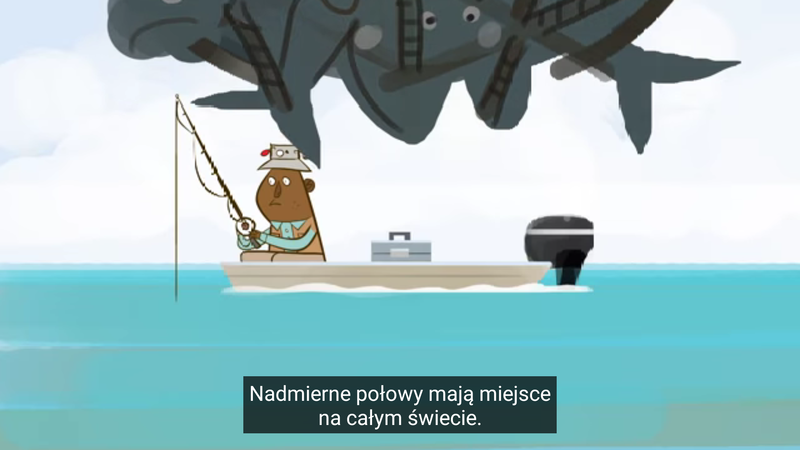}
   \caption{The video player screen with subtitles}
   \label{fig:Play_video}
\end{figure}

The application works as a regular player, so between choosing error categories the users can enjoy watching videos with subtitles (Fig. \ref{fig:Play_video}), as an edutainment activity \cite{rapeepisarn2006edutainment}.

\subsubsection {Error detection and category choice dialog}

Once the users spot a mistake in the subtitles appearing in the video mode they can click the middle round button on the remote (Fig. \ref{fig:mibox}) to pause the video. The application then overlays the error detection and category selection dialog over the video; on the right the subtitle is visible in the context of the subtitles surrounding it (Fig. \ref{fig:error_choice}). The user can then select if this is the subtitle they meant to pause at, and then they can choose and save the appropriate error category for the selected subtitle, at the same time restarting the video playback from a little before it was paused. 

\subsubsection {Error categories}
The error categories whose feasibility we decided to explore are based on key categories of errors in subtitles as extrapolated from OTPedia\footnote{The wiki site of TED Translators features subtitling guides to style, workflow and most common considerations for subtitles, including errors and formatting issues. The wiki can be found online at:  https://translations.ted.com/Portal:Main} (TED Translators' wiki with resources used as training and reference materials for translators). They are selected with the assumption of being easy to detect and interpret by viewers without much prior training.

\begin{itemize}
\item Timing: if the subtitle starts too early, or too late, or finishes too early in relation to the audio track, or if it disappears too quickly and it is impossible to read it - an important consideration when audio is present according to Armstrong \cite{armstrong2013development}.
\item Grammar: if there is a grammatical mistake: either with the tense, form, ending, punctuation or spelling.
\item Meaning: if there is difficulty understanding the meaning of the utterance or a suspected difference in meaning, especially in translation.
\item Style: if the wrong register is used in relation to the topic of the talk, or if the wording of a phrase is awkward, a potential calque, or could be corrected.
\end{itemize}

\begin{figure}
\centering
   \includegraphics[width=0.5\textwidth]{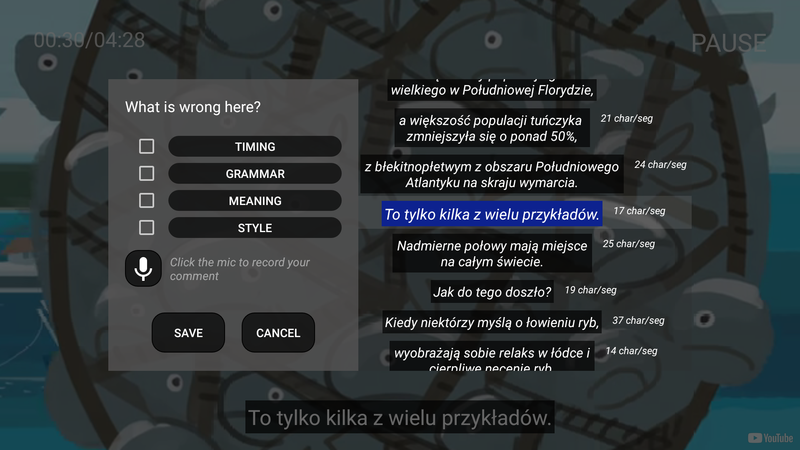}
   \caption{The advanced dialog allows users not only to select error categories, but also to record voice comments and see the subtitle reading speed.}
   \label{fig:error_choice} 
\end{figure}

These four main error categories are also more intuitive, unlike existing models of quality assessment of subtitles by professionals \cite{romero2018quality}, and they focus on information which could be useful for reviewers to pinpoint the errors later. This is unlike existing studies of subtitle quality assessment that focus on error detection without the intent to make later corrections of the same texts easier, as they largely relate to live TV subtitling, e.g. by Ofcom\footnote{Ofcom is the communications regulator of the UK, operating under Communications Act 2003 to ensure great quality of communications services in the UK. More information on Ofcom can be found online at: https://www.ofcom.org.uk/}. It is worth noting, that the Android TV remote used in the study was equipped with a microphone that could be used to provide additional comments or hints for translators through an option available in the basic interface (called advanced) visible in Fig. \ref{fig:error_choice}. This interface also allows the users to choose either one main, or a few error types in each subtitle, and display the reading speed required for it. 

To carry out our tests, however, we have developed a simplified interface (called simple), which allows the users to mark only one error category. This was done to estimate if the participants would request more, or less complexity in the error detection dialog, and to keep the task middle-weight as a fusion of edutainment (postulated error detection only, without categorizations) and full correction crowdsourcing (existing basic interface allowing the users to record short messages, justifying their choices and offering corrections).

\subsubsection {Settings}

Apart from the choice between the simple and advanced interface mentioned above, it is possible to choose to work on videos based on the audio and subtitle language combinations, where choosing the same language for both would limit the videos to transcription tasks only. To aid in further study iterations we introduced the "testing mode" to limit the TEDx video selection to our predefined videos.

\begin{figure}
\centering
   \includegraphics[width=0.5\textwidth]{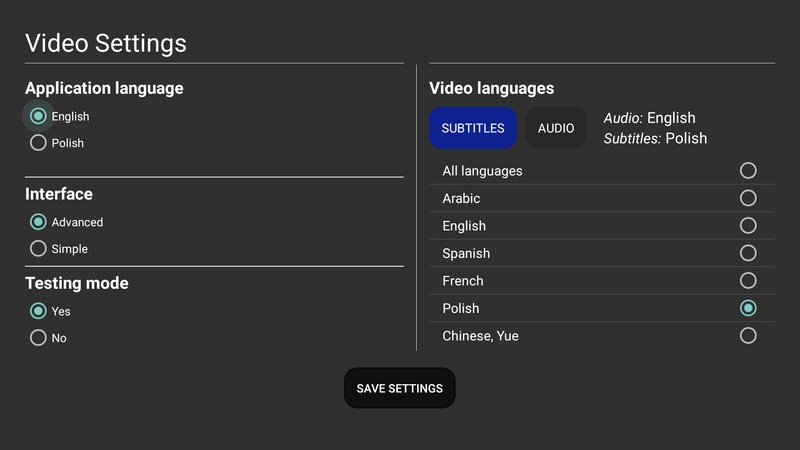}
   \caption{The settings allow the user to switch from the simple to the advanced mode, as well as limit the choice to specific videos in the testing mode.}
   \label{fig:settings} 
\end{figure}

\subsection {The journey of application development}

The application development process was based on early usability tests, co-design and consultations with the TED Translators community members. It was also informed by insights concerning empowerment and engagement from our aforementioned Living Lab's activities centered around older adults. Based on these we scaled down our initial idea of editing faulty subtitles to error detection, which is more aligned with the need for simplicity, the edutainment aspect, and more suitable for control via the remote. Since then, the development of the application has focused on the unique way of interaction with the app using the simplest remote, as well as creating a nonintrusive interface to promote the fun aspect of the task. In line with Android TV guidelines, we designed a pop-up dialog system that appears on screen only when the users press the middle PLAY/PAUSE button, as this way they get the full screen experience without any clutter elements that could distract them. Some of the other early major changes included: 

\begin{itemize}
\item Testing multiple combinations of the remote buttons.
\item UX color coding, sizing (readability).
\item UX flow and default behaviors (settings, auto-selection).
\item Interface labeling and action defining for clarity.
\item Changes to initial error categories (made the same for translation and transcription).
\item Overlay view for subtitle context visibility.
\item Subtitle skipping forward/backwards functionality (timing adjustment for the needs of older adults and consistency).
\end{itemize}
While the application was received favorably in our early tests, we decided to field test it using a structured protocol with older adults in the context that it was meant to be used (Living Lab): their own TV sets at home. 

\subsection{Study design, participants and testing protocol}
\begin{figure}
\centering
   \includegraphics[width=0.5\textwidth]{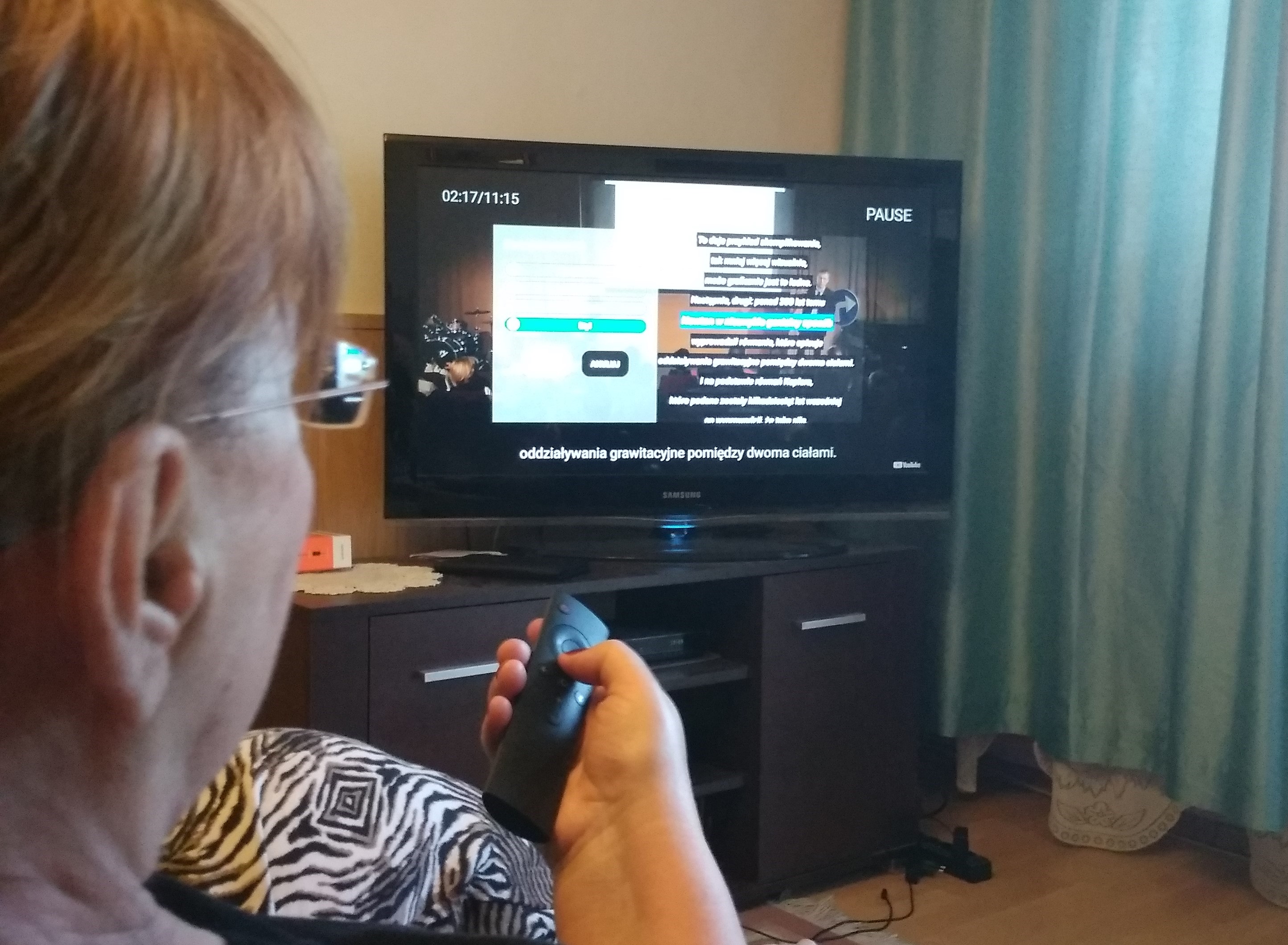}
   \caption{The at-home setup with a TV set facilitates the edutainment value of the task.}
   \label{fig:smarttv} 
\end{figure}

As it was mentioned above, we decided to employ the distributed Living Lab approach. In particular, we invited older adults to participate in the pilot study supervised by the researchers.
The research protocol, involving individual testing at home (Fig. \ref{fig:smarttv}), consists of the following main elements: an introduction to the study, the DigComp survey testing familiarity with computers and variety of ICT tasks, a semi-structured introductory interview that tests openness to new experiences as well as familiarity and preferences regarding subtitles, the explanation of the main elements of the project, that is the study and its benefits, an introduction to subtitles, and an on-paper exercise training error category detection skills, a demonstration and a hands-on test and the free interaction with the application and our five pre-selected and redacted test videos. The whole protocol takes about two hours to complete, and the reported preliminary tests were conducted in the first months of 2018.

\subsubsection{Introduction}

As the first step, we thanked the recruited participants for wanting to take part in this research and informed them that they can stop at any time and for any reason. We read the declaration of consent to them and asked them if they would like to sign it, and only then continued the research session.

\subsubsection{DigComp based survey}

The survey is a tool for measuring indicators of digital competence, developed by IPTS, and funded by the European Commission. It uses questions related to the Digital Skills
Indicator\cite{digcomp2014indicators} and broad ICT competence areas related to information (e.g. reading online news), communication (e.g. sending e-mails or making video calls), content creation (e.g. creating content, programming), safety (e.g. using anti-virus or a firewall) and problem solving (e.g. installing new devices, using Internet banking), based on the Digital Competence Framework \cite{ferrari2013digcomp}.

\subsubsection{Introductory interview}

The semi-structured in depth introductory interview aimed to gauge two main issues. First, familiarity and attitude towards subtitles with questions related to cinema and viewing habits. Second, openness to new experiences, including questions about habits related to new experiences, especially educational, and key interests and topic preferences.

\subsubsection{On-paper exercises}
The paper exercises are a key component of our step-by-step introduction method for user empowerment as they allow the older adults to become familiar with key notions in subtitling. This includes an explanation of what subtitles are for, and who can benefit from them, as well as some technical information such as reading speed, maximum line length and best places to introduce line breaks, and finally the specific task of subtitle error categorization. 
The exercises consist of a selection of text snippets from our chosen test videos, each containing either one, or two errors. Towards the end of the task, the results are discussed with the participant to ensure that they understand the key principles of subtitling, as well as the error categories in the application. The test also allowed us to estimate the confidence of older adults when detecting language errors.

\subsubsection{Using the DreamTV application with test videos}

For our study we have selected five videos to represent different types of challenges. For this reason, these videos were controlled for the following characteristics:

\textbf{General features}:

\begin{itemize}
\item a) Topic.
\item b) Video length.
\item c) Source language (spoken).
\end{itemize}

\textbf{Ease of comprehension}:

\begin{itemize}
\item d) Jargon saturation.
\item e) Required reading speed expressed in characters per second (ch/s).
\item f) Presence of the speaker on-screen.
\end{itemize}

\textbf{Types of errors}:

\begin{itemize}
\item g) Error source (machine or human).
\item h) Error saturation.
\item i) Error category.
\end{itemize}

\begin{table}
\caption{Characteristics of the videos used in the study. (Explanations: (a) Topic.
b) Video length.
c) Source language (spoken).
d) Jargon saturation.
e) Required reading speed expressed in characters per second (ch/s).
f) Presence of the speaker on-screen.
g) Error source (machine or human).
h) Error saturation.
i) Error category.)}
\label{videos}
\begin{tabular}{lp{4cm}p{8cm}}
	\hline
	\textbf{No} & \textbf{Video} & \textbf{Characteristics} \\
	\hline
     1 & \textit{Rebuilding Dreams One Bedroom at a time} by Joanna McCoy at TEDxKraków & a) Activism b) 10 min. c) \textbf{Polish} d) moderate modern e) medium (around 10--18 ch/s) f) on-screen g) human natural and simulated h) high-moderate (around 1 error /40 sec of video) i) mostly grammar with some timing \\
    \hline
    2 & \textit{General Einstein Theory of Relativity} by Krzysztof Meissner at TEDxMarsza\l{}kowska & a) Physics b) 11 min. c) \textbf{Polish} d) moderate science e) medium (around 11--18 ch/s) f) on-screen g) human natural and simulated h)  moderate (around 1 error /1 minute of video) i) predominantly grammar \\
	\hline
   3 & \textit{Will the Ocean ever run out of Fish} by Ayana Elizabeth-Johnson and Jeniffer Jacquet & a) Environment b) 10 min. c) \textbf{English} d) high e) very fast (around 16--21 ch/s) f) off-screen narration over a cartoon g) human natural h) very low (around 1 error /2 minutes of video) i) mixed \\
   	\hline
  4 & \textit{The hidden ways stairs shape your life} by David Rockwell & a) Culture b) 3 min. c) \textbf{English}  d) low  e) fast (around 13--21 ch/s) f) on-screen and off-screen g) Machine Translation with no corrections other than line alignment h) very high (around 1 error /10 seconds of video) i) mostly style and grammar \\
   	\hline
  5 & \textit{Inventing is the easy part. Marketing takes work} by Daniel Schnitzer at TEDxPittsburgh & a) Travel and technology b) 5 min. 30 s. c) \textbf{English} d) low e) medium (between 11--18 ch/s) f) on-screen g) MT with human corrections for obvious mistakes h) high (around 1 error /30 sec. of video) i) mostly grammar, style and meaning \\
   	\hline
\end{tabular} 
\end{table}

The videos selected for the preliminary study, visible in Table \ref{videos}, allowed us to observe a variety of factors at play, in order to determine the most interesting areas of further inquiry. The machine translations were generated using the original human-made transcripts, which were then imported to SubtitleEdit\footnote{SubtitleEdit is a free open-source subtitle editor with a wealth of useful features which available online at: http://www.nikse.dk/subtitleedit/}. This software has the option to use the Google Translate MT engine to generate subtitle translations; the generated lines then had their alignment fixed to reflect the transcript. The human natural errors are a combination of organic errors from the community, as well as errors introduced by researchers based on the most common translation/transcription errors lists on OTPedia\footnote{The TED Translators wiki features common error lists for many of the project's languages. To see the list for Polish refer to the materials available online at: https://translations.ted.com/Polish} to introduce varying error saturation levels. The tests were done with Polish language subtitles, as they were conducted with older adults in Poland.

\subsubsection{Exit IDIs}

In the exit interviews, we asked, among other things, about the feelings towards the general task of finding mistakes in subtitles, and in particular about their attitudes towards different videos we have tested. We also asked for any suggestions for increasing the attractiveness of this task in general, as well as concerning the topics of the videos, the reading speed, the preparatory exercises and the controls (remote) and setup. Additionally, we asked the participants if they would like to continue this task on their own, if they enjoyed the language part of the task and if they would like to try more/different types of language-related tasks in the future.

\section{Results}

Below we present the characteristics of the participants with the study summary, followed by some more detailed results.

\subsection{Study group and research highlights}

We invited seven older adults to our preliminary exploratory study: three female participants and four male participants. The study group was thoroughly selected from our Living Lab in Warsaw to cover several conditions: different age, occupation (one active person before retirement, two retired but still professionally active, two retired for several years and two retired for more than fifteen years). All participants live in Warsaw, the capital city of Poland, and they are native Polish speakers with limited or very limited English skills. There was a 20 year age span: the youngest participant was 60 years old and the oldest one was 79, with a mean age of 70,85 (SD=6,87). 

Based on introductory interviews and DigComp surveys we can describe this group as active users, above basic ICT skills. Everybody, except one, has smartphones, and the majority have Smart TVs. They are rather intensive and frequent Internet users, more than once a week.

Older adults in our study, even ones who report using the Internet only occasionally, and owning no smartphone, are very comfortable using Smart TVs at home. They do not see them as complex, but rather "interesting" (P4) and "useful" (P3), and "quite easy after some practicing" (P1, P2), some picked up on how to navigate them with a remote control within minutes (P1, P2, P3, P5), and all of them (P1--7) were able to learn how to perform the task in the DreamTV application in just one session. In case of P6 and P7, the learning curve was lower because their remote was similar to the one used in the study.

While all of the participants report that they rarely watch movies with subtitles, and prefer the voice over (P1--7) or dubbed content (P6), they see subtitles as "useful" (P5, P6, P7), as "cheaper and faster to make than voice over" (P2), and the task of improving them as "pleasurable and fun" (P3), "quite pleasant" (P1, P2) and "manageable" (P5). Most appreciate how "interesting" the videos are (P1, P3, P4, P6, P7), but one participant wished the topics were more "practical and useful for me" and said it was "not their thing" (P5). While the older adults were either good or very good at finding stylistic mistakes related to their natural language experience, especially in MT texts, they had trouble with technicalities such as timing, and small grammatical mistakes, like punctuation. Additionally, all of them complained about the reading speed required to understand the subtitles and sometimes paused the video to read the previous lines (P3, P4, P5, P6). While very few of the lines were above the 21 ch/s, which according to TED Translators' guidelines is the maximum reading speed, even the common industry standard of 17ch/s maximum was an issue. 

\subsection{Detailed results}

\subsubsection{On-paper exercises}

The participants were very sure of their answers when taking the tests, and most of them were able to explain their reasoning for choosing one error category over another, or marking a few at once. 
However, there appeared to be a dilemma relating to ambiguity and an overlap in the categories perceived by the participants, which is typical, as one type of error can be easily related to others. For this reason, we provided the option to mark multiple errors in the basic interface of the application.

The biggest issue was with the subtitle stylistic conventions, such as correct subtitle division (e.g. not leaving linking words at the end of the line and breaking up syntactic wholes), or the written format of sound information.

\subsubsection{Test videos}

The participants were mostly very confident when marking mistakes, and in the videos which were more saturated with grammatical and stylistic mistakes (videos no. 2, 5), they detected many of them. The stylistic aspects connected to language conventions were easy to pick up for older adults, especially in video no. 5. However, in video no. 4, which had machine translated subtitles, the stylistic problems were so saturated, that the meaning of large parts was unclear (P6, P7) or the whole video was unclear to some study participants (P1, P2, P4), even when reading the dialog list (Fig. \ref{fig:error_choice}). Moreover, video no. 4, which was Machine Translated, did not have any compression done to the subtitles, which pushed the reading speed up. The dialog list visible on the subtitle error choice screen (Fig. \ref{fig:error_choice}) proved to be very useful for two reasons. The main reason is that it was common that all of the study participants (P1--7) paused the video too late, and they had to use the dialog list to go back to the subtitle with the error, even reading it out loud (P3, P1, P6). Additionally, it helped comprehension as when the study participants felt they were getting lost, because of high reading speed required, or the jargon, they would pause the video and read the subtitles to make sure they understood the text (P3, P4). One surprise was that none of the study participants detected any timing errors in the videos. When asked about this, one participant replied that "they had to read" (P5).

While some study participants were indecisive about choosing the main error category (P4, P5) others (P3) were so eager to continue that they kept forgetting to mark it, naturally wanting to mark an error only. This correlated to how engaged they were in the topic of the video. The more interesting they found the video, the fewer mistakes they detected (P1--P4, P6).
Some participants requested to introduce an in-app tutorial, which would allow them to mark errors "without the worry that they will make a mistake" (P1, P4). Some study participants found the reading aspect to be difficult because of impaired vision. However, they just sat closer to the TV set to mitigate the effect and continue the task (P1, P3), but for most this was not an issue.

\subsubsection{Exit IDIs}

The participants felt they were capable of detecting errors in Polish subtitles for the videos (P1--7), and one of them even said: "There should be more subtitle testers like me, but not young people because they have little experience" (P3). Most of them claimed that they would be very happy to be able to do the task for a longer time by themselves with new videos like these (P1--4), but all of them would welcome the incentive of choosing topics interesting for them (P1--P7), as well as controlling for time ("The movies should be shorter, then I could watch anything! Just give me ten 5 minute films and I can do that for an hour" P3). Also, some of them (P2) are willing to do this activity in intergenerational context, i.e. with their grandchildren, who live abroad, when they come home to visit them during the holidays. Moreover, the videos in the study were deemed as quite interesting, and most of the study participants said that they "learned a lot" (P1--4). However, the participants had some issues with the specialized jargon used by the speakers (e.g. "it is not explained what is this photon" or "Spiderman, this is not Polish" by P2 and "kryptonite, must be a mistake" by P5, P6) and with the subtitles disappearing "too fast" (P1--7). While they enjoy finding errors in subtitles, they would not be interested to create their own, as it is "too much typing and time" (P1, P2, P4), apart from one who said that they "would have to try" (P3). They also said they enjoyed the language aspect, and they would be willing to try other language-related tasks in the future (P1--3, P6, P7).

\subsection{Notes on errors marked}

None of the synchronization errors were found, but also none were incorrectly marked. Stylistic errors were marked as such if the problem was with language style, while largely ignored when it was connected to subtitle conventions and formatting. The most common errors found were thus stylistic and grammatical errors related to spelling and general language correctness. Capitalization and punctuation errors, belonging to the grammar category, were difficult to spot for the older adults, we expect because some of the errors required them to remember the punctuation from the previous lines. On top of this, many study participants interpreted the "meaning" category, as something to mark when they encountered an unfamiliar term, or a divergence between spoken phrase and its transcription for the native language, which made for a few false positives to mark more specialized words and subtitles with higher compression rates. 

\section{Discussion}

Based on the exploratory research and pilot evaluation of our setup of online crowdsourcing translation process support via Smart TV platform we identified five promising areas of consideration when designing similar systems. These include (1) autonomy and freedom of choice, (2) physical comfort and (3) cognitive comfort to aid comprehension as well as (4) building confidence and (5) providing the edutainment value. While these five considerations form a starting point as preliminary guidelines for further inquiries, it is important to remember the limitations of this exploratory study, as it was conducted with older adults in Poland, among the more active and ICT-aware well-educated individuals belonging to the economic lower and middle-class\footnote{The best status class to describe this group would be Intelligentsia, which is the class of educated, but not necessarily well-off people in Poland.}. The following discussion is aimed to inform the design of systems which could better include such older adults in ICT-dependent crowdsourcing tasks. For each of the signaled general considerations there are preliminary proposals of best practices to mitigate or overcome the observed barriers and accommodate preferences which are discussed below. 

\subsection*{Promising considerations for designing crowdsourcing systems for older adults:}

\subsection {Autonomy and freedom of choice}

The question of the ability to choose the topic, duration and engagement level with the content at any time appeared to be of crucial importance for the engagement of our group of older adults.

\subsubsection{Freedom to choose the topic} 
The older adults wanted to \textbf{freely choose the content to watch}, based on their preferences. This is of key importance, as in our use case their primary motivation comes from the alignment with their interests providing the edutainment value.
Thus, one crucial insight is that it is necessary to introduce older adults' profiling, perhaps in the form of a few questions during the initial setup of the app to allow the participants to be served videos based on their interests, as well as topic categorizations within the app. 
Moreover, for longer studies it is important to introduce the function to search by the topic of the video, as there are clearly some topics favored by older adults and because in this case their interests act as a motivational element. This could be supported by a user profiling system coupled with an adaptive mechanism, that monitors and follows user preferences and their changes over time.

\subsubsection{Freedom to control task duration}
The older adults wished to be able to choose how long they want the videos to be, and how much time they would like to devote to this task overall, which is another potential adaptive service.
\subsubsection{Freedom to contribute as little or as much as one wishes}
The older adults were pleased about the learning potential of the videos, but they were also eager to finish watching the videos early if they were not as interested in the topic (P1, P5). This suggests that the task is just as engaging as the video, which corresponds to the edutainment concept. However, the more engrossed older adults were the fewer errors they seemed to detect, which suggests prompts or some other technique could be introduced to remind the viewers of the task. This, combined with their unwillingness to create subtitles themselves, as it is too much time and effort, shows that \textbf{this type of crowdsourcing should rely on quantity of contributors, rather than their accuracy} because participants are unlikely to go back to the videos to watch them again and verify their own accuracy when detecting errors. 

\subsection {Physical comfort and familiarity}
It indeed seems that our group of older adults interacted with Smart TVs in a more familiar, and definitely more comfortable fashion than with computers due to their long-time familiarity with the setup and controls, as well as the physical comfort granted by their leisure home environment.

\subsubsection{Physically comfortable setup (TV)}
Regarding the Smart TV interface, \textbf{people who watch TV at home already have a very comfortable setup for spending time in front of the TV Set}, unlike their computer setup at home, which, as we observed, is mostly designed for short sessions. This corresponds with numerous market research done e.g. by Nielsen. Despite the task appearing in the same setup as one of their favorite leisure time activities, that is watching TV, our group of older adults viewed using the DreamTV application as more demanding, which raises the questions, how much time would they be willing to devote to it, and if the solution is sustainable in the long run. These ought to be further researched. 

\subsubsection{Ease of navigation with clearly labeled controls (remote)}
The \textbf{Android TV interface proved to be very easy to navigate}, as the older adults participating in our study are quite used to navigating the menus and teletext by clicking arrow buttons on the remote. We observed that this was a largely familiar and comfortable experience for them, despite the remote being much different from theirs for P1--P5. One suggestion we received was to mark the remote control with buttons of different colors, as now "they look the same" (P3, P4) or separate the direction keys (P6, P7). The colors could then correspond to the colors of actions which can be taken on the screen to assure cohesion of design between the physical controller (the remote) and the application design on screen.

\subsubsection{Ability to take breaks}
The older adults valued the task also for the ability to take breaks, pause, make tea (P3, P5) and readjust without losing sight of the context and feeling pressured into hurrying. This suggests that there ought to be no timeouts or time limits imposed in the design of other crowdsourcing tasks.

\subsubsection{Adjustable text size and colors}
There ought to be the possibility to adjust the size of the letters and introduce high contrast selection to accommodate vision impairments. The text size should be adjustable based on the possible needs of older adults, as not all of them have their TVs at a comfortable distance to read the fonts, and have varying degrees of light in the living room. While some of them compensated by moving the chairs closer to the screens, for long time use it would be better if they could stay on the couch where they are comfortable. We expect this to be a lesser issue in societies where subtitle use rates are higher.

\subsection {Cognitive comfort to aid comprehension}

\subsubsection{Adjustable reading speed}
\textbf{The required reading speed ought to be in the low range (14 ch/s at most)} or possibly the speed of the video playback could be adjustable, or even adaptive. What also follows, is that in order to allow older adults to be more effective at finding mistakes in Machine Translated subtitles the software has to use language compression algorithms, to cap the required reading speed at 14 ch/s for the ease of comprehension. This also holds true for human organic errors, as some human-translated videos, especially by untrained volunteers, use close to literal translation with no compression\footnote{More information on problems and techniques of subtitle compression can be found on TED Translators' wiki in the following location:https://translations.ted.com on the "How to Compress Subtitles" wiki page.}, which tends to make subtitles longer.

\subsubsection{Ability to pause, go back and view the context for reference}
It should be possible to \textbf{view the video transcript} and to go back to any previous point, as older adults have slower reflexes and often clicked pause too late. They also welcomed the ability to pause and check the context of the dialog list\footnote{More on this feature was explained in section 3.1.3, while the dialog list providing context is visible in Fig. \ref{fig:error_choice}.} for reference, in case they felt they may have missed something.

\subsection {Building confidence}
Based on our participatory observation and exit interviews, the older adults in our group consider themselves competent enough in their native language to detect errors, and with a well structured step-by-step tutorial, also errors in subtitles. With a more extensive tutorial on types of errors common in subtitles, combined with a quick refresher of grammar rules, especially focused on punctuation, the older adults will be a good group for this type of crowdsourcing.

\subsubsection{Making use of older adults' language skills}
Without extensive prior training and with some limited empowerment provided by the researchers, the older adults can \textbf{use their natural language experience to find mistakes}, especially with grammar and style, which works well with MT errors, but excludes technical problems such as formatting and synchronization. While making use of older adults' expert native user proficiency in language enabled crowdsourcing is very empowering, the question of the evolution of language also deserves attention. Some older adults were not familiar with technical or pop culture references (P2, P5 and P6) and they assumed these may have been mistakes. To mitigate this barrier we propose that a reference dictionary could be available to the participants via the app, in which they could verify their linguistic hunches and at the same time learn more modern terms for the contexts they encounter.

\subsubsection{Providing a theoretical and practical tutorial}

While grammar and style seem to be the best category for this type of task, the timing issues were not detected by anybody, neither in transcription nor in translation tasks. It is possible the difficulty with finding timing (sync) issues was due to not including this category in the exercises before the task, as they were made on-paper and were limited to the choice between "grammar", "style" and "meaning". Moreover, the technical aspects of the subtitling process, despite appearing in the protocol on-paper exercise, such as subtitle stylistic conventions, were largely overlooked, and would require more training, or an introduction of the subtitle-specific tutorial to the application. 
In general, it seems that tasks where the subtitles are a translation (videos no. 3, 4, 5) are a better target for language correction than transcription tasks (videos no. 1, 2) where older adults tended to try to focus if the words written are exactly as said when finding mistakes. (especially P5). We conclude that \textbf{it would be beneficial to introduce a tutorial within the app} that would prime the older adults to focus on the most common error types. 

\subsection {Providing the edutainment value}
The content we used for this activity, that is the TED and TEDx talks is quite engaging and broad in terms of domain. One challenge that exists then, is to \textbf{frame other crowdsourcing tasks in such way as to provide edutainment value} for the participants, which would allow them to consider this task to be closer to their leisure time than to work.

\subsubsection{Using wide-domain educational content}
As we observed the older adults from our group find improving the quality of subtitles for TED and TEDx talks to be an interesting and motivating task because of its educational value, and what has to be stressed, personally for them with the topics of their choice. Therefore it seemed that easily providing older adults with the choice of topic they are interested in, and at the same time ensuring that they are learning is a necessity when designing edutainment-like crowdsourcing. The DreamTV application facilitates this, as TED and TEDx talks are characterized by a very large domain, and multiple videos have a wider focus than just technology, entertainment and design.

\subsection*{Additional note}

As more ICT-proficient generations enter retirement age in Western societies, such studies may soon prove to be more generalizable, both because of rising levels of ICT literacy, and the introduction of affordable solutions such as the Android TV STB, like Xiaomi MiBox, which can turn any TV set into a Smart TV. As such, it is a promising setup for an increasingly larger group of older adults, who are more ICT-savvy and have access to cheap technology able to turn their TV sets into Smart TVs.

\section{Conclusions and further work}

\subsection{Overview}
The Smart TV setup and the interface are promising in the field of research with older adults, especially due to the familiarity, large screen size, general accessibility and comfortable at-home setup. Crowdsourcing subtitle errors is also perceived as an interesting task, which can be very engrossing and motivating depending on how it is framed. This study allowed us to evaluate the feasibility and direction of further inquiries in the area of TV-enabled crowdsourcing tasks, relying on native language proficiency of older adults. The key areas of interest when designing such systems based on our exploratory study seem to be (1) autonomy and freedom of choice, (2) physical comfort and familiarity, (3) cognitive comfort to aid comprehension as well as (4) building confidence and (5) introducing the edutainment value. Therefore, we consider our study an important voice, and a contribution into the discussion of vital research areas of human aspects of collaborative systems development and crowd work, raised in particular at CSCW in the recent years by numerous researchers, e.g. from Carnegie Mellon, Northwestern and Stanford Universities \cite{kittur2013future,culbertson2017have,zhu2014reviewing,yu2014comparison}. In particular we are convinced that the Smart TV interface is promising for developing sustainable crowdsourcing solutions, and platforms that could truly involve older adults in crowd tasks by lowering ICT barriers and motivating them in various ways, from providing engaging content and edutainment, to ensuring social
inclusion and enabling contribution to the society at large. 

\subsection{A follow-up long term study}
However, further work is necessary to verify our presented preliminary considerations in a follow-up long-term study. This is feasible, as we have promising findings that some participants would like to continue, and some (P2 and P3) in fact have already subscribed, with one participant (P2) suggesting to do the task with grandchildren, thus implying that the intergenerational dimension could be an additional motivator. Below we present a brief overview of five main areas of future research. 

\subsubsection{Sustainability of the solution and verification of preliminary findings}
Given our observations and feedback, in the long-term study we would like to verify the five above-mentioned design considerations as well as the sustainability of this solution in terms of: continued interest, likelihood of stopping playback to mark an error, detecting false error positives, as well as changing motivation depending on the context of the task (e.g. intergenerational). As observed in our preliminary study, older adults wish to engage in crowdsourcing if it resembles leisure activities they can do for fun. In general, we believe that many older adults will be able to sustain interest in the task, once it has a wider range of content available that may be better aligned with their interests.
\subsubsection{Error detection rates}
Another interesting research path is the examination of the inverse relationship between the study participants' interest in a specific video, and the number of errors they detect. In order to test this, we intend to allow older adults to add an option to rate each video they have watched. To further examine the differences between detection rates the long term study would feature videos with both errors naturally occurring after human translation, as well as originating from machine translation-aided tasks. As our results indicate in both of these scenarios, the quality of translations can benefit from the involvement of older adults as proofreaders marking style and comprehension problems as well as grammar issues. Such data could be later used to facilitate Machine Learning processes in the Machine Translation field, but also help improve the quality of human translations.
\subsubsection{Comprehension}
Yet another aspect of the further study could focus on ways of addressing the comprehension difficulties we encountered, caused by the need to focus on many aspects of the subtitles and making choices regarding the error categories, in particular, by removing error categories altogether, and just enabling an error to be reported. There exists one related ground of research: crowdsourcing the errors that are perceived as errors by the older adults, and comparing them to mistakes which are generally regarded by the translation communities and edutainment scene as a whole. Older adults often have signaled problems with the speed and comprehension of the subtitles themselves. Freely gathered data on perceived errors could better inform ways of making these forms more accessible to older adults.
\subsubsection{Ease of setup}
At the same time, although the Smart TV interface and our application proved to be enjoyable and easy to interact with, it is important to verify if older adults would be able to set up the system on their own, and if not, how to make it easier for them. This alone would pave way for further inquiries into the feasibility of using Smart TV services for older adults. 
\subsubsection{Motivation}
As is often the case with crowdsourcing, this task also depends on the scale of contributions. Even if some older adults, according to their psychological profile and motivation, may choose to just watch the videos for fun, the others may use the opportunity to detect errors, and with the effect of scale working the true errors would be detected more reliably. We would also like to test the motivational component involving statistics of the errors found, and possibly gamifying them. Once the DreamTV application has been more extensively tested and developed to include the suggestions, and to resolve the difficulties occurring, we would like to allow as many users as possible free access to the application, and to collect longitudinal quantitative data on the errors reported. At the same time, we would like to test the personal (increasing the positive self-image) and social (engaging in a community) potential of this type of task we postulated when designing the DreamTV crowdsourcing solution. We could test it by encouraging older adults to engage with the TED Translators community, members of which could request their help in proofreading their own translations with the help of the application.

\subsection{Closing remarks}
Although some of our findings can be generalized to inform future design further work needs to be conducted with larger and more diverse groups alongside follow-up long-term studies. For example, in terms of wealth, education, language abilities and proficiency with technology, as well as with younger adults or speakers of other languages (e.g. migrants). This will help to explore which of our preliminary findings may be universal, and which ones rather depend on factors specific to different groups. 

\begin{figure}
\centering
   \includegraphics[width=0.5\textwidth]{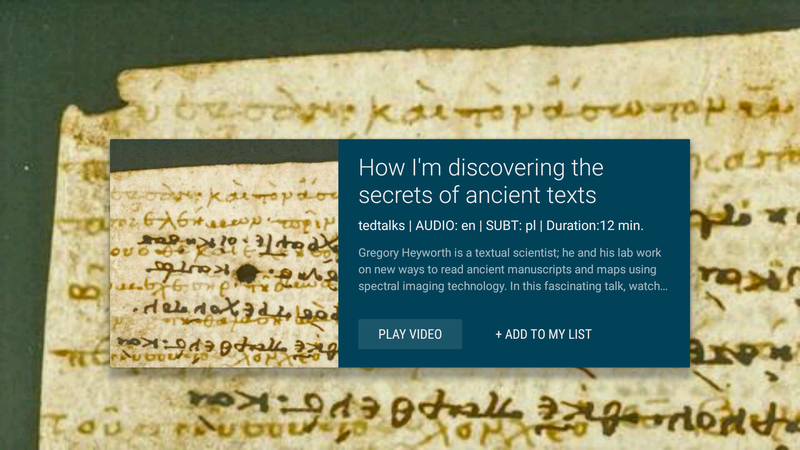}
   \caption{As the application allows users to select their favorite videos and save their progress it has potential for sustaining engagement, which is also helpful for conducting long term studies.}
   \label{fig:video_details} 
\end{figure}

\section{Acknowledgments}

We would like to thank the TED Translators community for their insights and work on making interesting ideas accessible to more people through their subtitles, as well as the older adults who participated in this study.
This project has received funding from the European Union's Horizon 2020 research and innovation programme under the Marie Sklodowska-Curie grant agreement No. 690962.

\bibliographystyle{ACM-Reference-Format}
\bibliography{bibliography} 

\end{document}